**Evaluating Trust in AI, Human, and Co-produced Feedback Among Undergraduate Students**

**Title Page**

- Title: Evaluating Trust in AI, Human, and Co-produced Feedback Among Undergraduate Students
- Author information:

    - Full names of all authors:
        - Audrey Zhang1 (ORCIDs: 1 0009-0002-8180-0714)
        - Yifei Gao2 (ORCIDs: 0009-0009-4571-1037)
        - Wannapon Suraworachet2 (ORCIDs: 0000-0003-3349-4185)
        - Tanya Nazaretsky3 (ORCIDs: 0000-0003-1343-0627)
        - Mutlu Cukurova2 (ORCIDs: 0000-0001-5843-4854)
    - Corresponding author's email: zhongyao.zhang.19@ucl.ac.uk
    - Author affiliations and countries:
        - 1 Department of Experimental Psychology, University College London, London, United Kingdom
        - 2 UCL Knowledge Lab, University College London, London, United Kingdom
        - 3 École Polytechnique Fédérale de Lausanne, Lausanne, Switzerland

## Abstract

As generative AI models, particularly large language models (LLMs), transformed educational feedback practices in higher education (HE) contexts, understanding students' perceptions of different sources of feedback became crucial for their effective implementation and adoption. This study addressed a critical gap by comparing undergraduate students' trust in LLM, human, and human-AI co-produced feedback in their authentic HE context. More specifically, through a within-subject experimental design involving 91 participants, we investigated factors that predict students' ability to distinguish between feedback types, their perceptions of feedback quality, and potential biases related to the source of feedback. Findings revealed that when the source was blinded, students generally preferred AI and co-produced feedback over human feedback regarding perceived usefulness and objectivity. However, they presented a strong bias against AI when the source of feedback was disclosed. In addition, only AI feedback suffered a decline in perceived genuineness when feedback sources were revealed, while co-produced feedback maintained its positive perception. Educational AI experience improved students' ability to identify LLM-generated feedback and increased their trust in all types of feedback. More years of students' experience using AI for general purposes were associated with lower perceived usefulness and credibility of feedback. These insights offered substantial evidence of the importance of source credibility and the need to enhance both feedback literacy and AI literacy to mitigate bias in student perceptions for AI-generated feedback to be adopted and impact education.

## 1. Introduction

Constructive and personalized feedback plays a crucial role in supporting students' learning, as it fosters confidence in their ability to improve and make progress (Ruijten-Dodoiu et al., 2025). Studies emphasize the importance of formative feedback for effective learning and its significant effects on student learning and satisfaction (Bandiera et al., 2015; Cavalcanti et al., 2021; Hattie & Timperley, 2007; Henderson et al., 2019; Hooda et al., 2022; Zhang et al., 2024). Due to the demands for feedback and high workload, instructors often struggle to provide detailed and constructive feedback to large student cohorts in universities (Dai et al., 2024; Henderson, Ryan, & Phillips, 2019). The teacher-student imbalance, combined with diverse student backgrounds and varying knowledge levels, further compounds this challenge (Carless & Winstone, 2020; Dai et al., 2024; Nazaretsky et al., 2024a).

The rapid development of Generative Artificial Intelligence (GenAI) tools, especially Large Language Models (LLMs), introduces new opportunities for enhancing the efficiency and effectiveness of student feedback provision (Cavalcanti et al., 2021; Dai et al., 2023). As LLMs are pre-trained on a large corpus of data, enabling them to learn complex human language, they demonstrate the capability to generate human-like content (Yan et al., 2023), including feedback generation. Recent studies confirm LLMs' potential to create high-quality formative feedback comparable to human expert feedback in various domains, including but not limited to data science (Dai et al., 2023), argumentative writing (Escalante et al., 2023; Steiss et al., 2024;), and programming (Gabbay & Cohen, 2024; Pankiewicz & Baker, 2023).

Despite these benefits, significant issues remain regarding LLM-powered AI feedback, including implicit and explicit biases embedded in the generated feedback content (Nazaretsky et al., 2025a), reliability and validity of feedback content (Chang et al., 2024), as well as broader ecosystem-level issues, such as the potential de-professionalisation of teachers due to potential cognitive atrophy at key pedagogical tasks like feedback generation (Cukurova, 2024). Perceived bias and trust in the AI feedback are key factors determining the adoption of AI feedback beyond its technical accuracy (Chan & Hu, 2023; Cukurova, 2024). For instance, a study by Cukurova et al. (2020) demonstrates that the perceived credibility of educational evidence is influenced by its framing, specifically lower perceived credibility in AI-framed evidence compared to traditional educational psychology and neuroscience frames. This suggests a potential bias depending on how feedback is delivered, particularly whether students should be informed about its source or can infer it from the content.

Including humans (in this case, the teachers) in the loop of AI feedback production could be a potential strategy to mitigate such AI bias (Nazaretsky et al., 2025a; Rüdian et al., 2025). The co-produced approach could potentially leverage complementary strengths, such as human intelligence enabling higher-level reasoning and criticality aligned with pedagogical goals (Luo et al., 2025; Pahi et al., 2024), contextual accuracy (Jamie et al., 2025), and socioemotional considerations (Er et al., 2024), while LLMs contribute by generating readable, consistent, and detailed feedback at scale (Escalante et al., 2023; Dai et al., 2023; Dai et al., 2024).

While previous studies have evaluated AI-generated formative feedback primarily on efficiency and accuracy metrics (Dai et al., 2023; Hao et al., 2021), research gaps remain in examining students' social-emotional perceptions and subjective responses to AI-generated feedback (Er et al., 2024; Nazaretsky et al., 2024a, 2025a). Students' reactions to feedback inform instructors about the effectiveness of their teaching strategies, enabling them to adapt their approaches to meet learning needs. A recent study compared students' perceptions of AI-generated versus human-created feedback (Nazaretsky et al., 2024a, 2025a), but did not involve co-produced feedback combining human and AI strengths. This study addresses these gaps by introducing a co-produced feedback approach in a HE context, comparing student perceptions across AI-generated, human-created, and co-produced feedback.

## 2. Literature Review

### 2.1. The Role of Feedback Providers in HE

#### 2.1.1. Human-generated Feedback

Human feedback generated by instructors in a HE context is the most common way to enhance student academic performance (Ajjawi & Boud, 2018; Gan et al., 2021). The effectiveness of human-generated feedback lies in instructors' ability to provide contextually relevant, personalised guidance that addresses individual student needs and learning trajectories. Its quality is thereby linked to instructor competence and knowledge. Effective feedback requires a deep understanding of learning objectives, assessment criteria, and pedagogical strategies for communicating complex concepts (Winstone & Carless, 2019). Instructors must possess both disciplinary expertise and feedback literacy to identify student misconceptions, provide constructive guidance, and scaffold the learning process.

However, scaling human-generated feedback presents significant challenges in contemporary HE contexts. Providing personalized, detailed formative feedback is inherently time-intensive and demands substantial instructor effort (Ajjawi & Boud, 2018; Gan et al., 2021). The task is further complicated by increasing student enrollments and persistent instructor shortages, resulting in unsustainable instructor-to-student ratios (Ruwe & Mayweg-Paus, 2023). These constraints create excessive workloads for instructors in large cohorts, despite the recognized importance and effectiveness of providing quality feedback (Er et al., 2024; Zhang et al., 2024). Given these scalability limitations, AI-generated feedback emerges as a potential solution to mitigate challenges associated with human-generated feedback (Lee & Moore, 2024).

### 2.1.2. LLM-powered AI feedback

LLM-powered AI feedback refers to feedback generated by advanced artificial intelligence models, specifically Large Language Models (LLMs), such as ChatGPT (https://chatgpt.com) and Claude (https://claude.ai). These models have been trained on vast data corpora to recognise complex linguistic patterns (Yan et al., 2023) and are capable of interpreting and generating human language with high fluency (Escalante et al., 2023; Ruwe & Mayweg-Paus, 2024). With these automating capabilities of LLMs, they offer opportunities for educational researchers to employ LLMs in various educational aspects, including generating questions (Wang et al, 2022), scoring essays (Seßler et al, 2025), and providing feedback (Cavalcanti et al., 2021). AI feedback has increasingly been explored for feedback automation in HE and shown promises across subjects such as language teaching (Nguyen et al., 2024; Zou et al., 2025) and STEM learning (Bewersdorff et al., 2023; Dai et al, 2023). Beyond reducing instructors' workload in producing feedback, it offers instructors opportunities to engage in more complex pedagogical tasks as well as improving students' learning outcomes through supportive and interactive environments (Lee & Moore, 2024). These models demonstrate comparable feedback quality to well-trained human evaluators (Cavalcanti et al., 2021), also providing more specific guidance for student learning (Nazaretsky et al., 2024b). In some cases, AI feedback is more readable and detailed than human feedback (Dai et al., 2023). For instance, McLaughlin et al. (2025) found that 85% of HE participants perceived GenAI feedback as valuable and helpful in improving their learning plans. Not only did it show no negative impact on students' learning outcomes (Escalante et al., 2023), but it also contributed to improving outcomes by fostering both cognitive development, such as writing ability (Hu et al., 2023; Neo, 2022), and metacognitive enhancement (Lee et al., 2022). These examples suggest that AI feedback has strong potential compared to human instructor feedback regarding quality and efficiency.

To control for the quality of AI feedback, ensuring it is factually correct (Elsayed, 2024) and valuable to students' learning (Lee and Moore, 2024) is essential. While LLM output accuracy cannot be fully controlled due to its stochastic nature, it can be partially structured through prompt engineering (Strobelt et al., 2022). This implies the accuracy of AI feedback relies on precise, well-crafted prompts, as vague or broad prompts can lead to inaccurate or unhelpful responses (Escalante et al., 2023), which warrants further experimentation, especially when deploying LLMs in real-world learning contexts. Moreover, this prompt design and evaluation task imposes a workload and should be considered in discussions of AI-generated feedback implemented by human tutors.

In addition to content quality issues, LLMs face technical and organisational limitations, including hallucinations, bias, reliability concerns, accessibility issues, and privacy concerns (Suraworachet et al., 2024). While valid, these issues fall outside the study's scope due to technical considerations. Instead, this paper focuses on students' characteristics and perceptions of feedback quality and sources, which affect their trust in AI, as expanded in Section 2.2.

#### 2.1.3. Co-produced Feedback

Beyond independent human or AI capabilities, there has been a growing interest in examining and building joint capabilities combining human-AI intelligence. Hybrid intelligence, defined briefly as the synergistic combination of human and artificial intelligence, could lead to emergent intelligence greater than the sum of the individual agents' intelligence, or at least greater than the maximum of human or AI performance on a given task (Cukurova, 2019, 2024). Such hybrid approaches, which aim to complement or augment human intelligence with AI, have been increasingly and conceptually proposed in educational contexts. (e.g., Järvelä et al., 2023; Molenaar, 2022; Holstein et al., 2020). Though a systematic review has yet to consolidate these hybrid approaches specialised in educational context, several empirical studies have investigated its potential, for example, human-AI collaboration to identify students' difficulties in solving STEM tasks (Cohn et al., 2024), to recommend targeted computer science assignments for students (Pereira et al., 2023) or to support argumentative writing revisions (Luo et al., 2025).

Aligning with the hybrid approach, human-AI collaboration in feedback provision could combine both agents' strengths and mitigate their weaknesses through co-produced feedback. Human-AI co-produced feedback refers to the process of generating and delivering feedback that is jointly produced by humans and AI through a series of iterations. It can take various approaches: AI-initiated feedback reviewed by humans, human-generated feedback refined by AI, or an iterative improvement of both approaches (Jamie et al., 2025; Luo et al., 2025; Pahi et al., 2024). For example, Luo et al. (2025) deployed AI feedback based on a text analysis tool that incorporates linguistic features, in combination with teacher feedback on students' argumentative writing drafts. Through a comparative analysis of the drafts, it was observed that AI targets lower writing level enhancement, which is particularly constrained by the AI tool, compared to teachers' focus on higher writing levels. Pahi et al. (2024) investigated the quality and timeliness of teacher (teaching assistant, TA) and AI feedback. Their co-produced feedback conditions included independent AI and TA feedback generation, AI feedback reviewed by TAs, and AI-improved TA feedback. Distinctive benefits were found: TAs excel at identifying technical problems, while AI provides effective motivational support and scalable examples. Moreover, the co-produced approach enhanced feedback clarity and accuracy, suggesting augmented human-AI capabilities beyond individual performance. However, Pahi et al. (2024) focused solely on students' experience with TA support, without comparing perceptions across feedback types, leaving gaps in the feedback type.

The co-produced approach tends to focus on studying AI-generated feedback reviewed by teachers, citing teachers' expertise in maintaining pedagogical focus and the scalability of AI. While this may seem justified, it may not fully account for the situations where teachers, such as novice TAs, have limited experience in teaching and managing large university classes. This suggests a need to consider how AI is utilized in real-world HE practice, particularly in the context of TAs co-producing feedback with AI, which serves as a language tool for proofreading feedback.

Other challenges, including communication difficulties, trust issues, ethical dilemmas, and coordination complexities, can compromise the effectiveness of hybrid systems (Bansal et al., 2021; Buçinca et al., 2020; Zhang et al., 2020), which should also be taken into consideration.

### 2.2. Factors Impacting Student Perceptions of Feedback

Student perceptions are important to explore as they influence attitudes and behavioural intentions toward adopting automated feedback systems (Obenza et al., 2024; Ramirez et al., 2024; Shamsi et al., 2022). Prior studies assess students' judgment accuracy by benchmarking them against expert evaluations (Andrade, 2019; Falchikov & Goldfinch, 2000). However, applying expert evaluation frameworks directly to students creates challenges, as students are learners rather than professionally trained experts. Students often have limited specialisation and may not fully grasp the broader educational objectives of formative feedback, potentially misaligning with instructors' intentions (Gibbs & Simpson, 2005; Tai et al., 2017).

Student perceptions of feedback can be measured using several evaluation frameworks, depending on the target dimensions. To enhance evaluation accessibility, it is crucial to employ frameworks that are easy for students to understand and that facilitate the measurement of their perceptions toward feedback sources, while still preserving fundamental evaluation principles. For example, Ruwe and Mayweg-Paus (2023) used the Feedback Perceptions Questionnaire (FPQ) scale to evaluate students' perceptions of feedback message quality, specifically focusing on its usefulness and fairness. Moreover, they utilised the Muenster Epistemic Trustworthiness Inventory (METI) scale (Hendriks et al., 2015) to assess students' perceptions of the feedback provider's competency in expertise, integrity, and benevolence. Er et al. (2024) assessed students' perceptions based on four key dimensions: fairness, usefulness, development, and encouragement, on instructor vs AI feedback.

Previous research highlights key factors shaping learners' willingness to accept feedback, such as objectivity (Roberson & Stewart, 2006), usefulness (Henderson et al., 2019; Ryan et al., 2019), genuineness (Hirunyasiri et al., 2023; Shute, 2008), and credibility (Roberson & Stewart, 2006). Regarding objectivity, informational justice theories suggest feedback should be perceived as fair and accurately reflecting the recipient's performance (Boud & Molloy, 2013; Roberson & Stewart, 2006). Besides, effective feedback should be specific and directly related to the learner's response (Shute, 2008). For learners to engage with feedback regarding its usefulness, it should provide relevant and applicable information directly related to the recipient's context (Ryan et al., 2019). Regarding genuineness, feedback should maintain a genuine, encouraging tone (Er et al., 2024; Hattie & Timperley, 2007), reflecting authenticity and sincerity (Ruijten-Dodoiu et al., 2025). It should be grounded in honesty and earned, avoiding overuse (Hirunyasiri et al., 2023). Credibility requires that learners perceive feedback providers as secure and ethical sources. Trustworthiness and reliability are key safety aspects, while ethics includes responsibility, privacy, and freedom from biases (Yan et al., 2023). Driven by a review of the literature and aligned with recent studies measuring students' perceptions of AI-generated feedback (e.g., Nazaretsky et al., 2024a, 2024b, 2025a), this study investigates students' perceived *usefulness, objectivity, genuineness,* source *credibility* of feedback, and general trust in AI-edTech.

### 2.3. Students' AI Feedback Biases

Students' perception biases toward AI feedback as its providers substantially shape how users evaluate AI-generated output primarily stem from two directions: algorithm appreciation (Logg et al., 2019) and algorithm aversion (Dietvorst et al., 2014). Algorithm appreciation may lead to overtrust in AI, driven by overly optimistic expectations of capabilities. It is often caused by limited AI literacy, leading to automation bias, where users assume AI feedback is inherently more accurate than human output. For example, Ruwe and Mayweg-Paus (2023) found that AI was perceived as the most trustworthy source of expertise, even more so than educators or peers.

However, algorithm aversion emerges when students doubt AI's capacity to provide credible, accurate, or tailored feedback to student learning needs. This aversion can potentially diminish the perceived efficacy of technological capabilities and the pedagogical impact of AI systems (Le et al., 2025). Such scepticism may stem from distrust of AI systems, the belief that AI cannot fully grasp the nuances of human learning contexts, or discomfort with the impersonality of machine-generated responses. For example, Abendschein et al. (2024) found that although a fictional AI feedback system was viewed as competent and trustworthy, it received lower evaluations than human graders providing identical feedback. These conflicting findings reinforced existing biases in providing feedback sources and underscore the importance of examining biases and trust in AI and human-generated feedback, as these factors significantly influence student interactions and their willingness to rely on both systems. Likewise, students may view AI feedback as lacking credibility or genuineness compared to human feedback, potentially leading to more negative evaluations, regardless of the actual quality (Nazaretsky et al., 2024a, 2024b). These contrasting findings underscore the challenge of evaluating AI-generated feedback, as biases vary significantly across diverse student groups and contexts.

Underpinning these biases is trust, which emerges as one of the most crucial factors influencing students' decisions when choosing between human- or AI feedback (Peng & Wan, 2023). Specifically, various studies have indicated that trust is a significant predictor of both attitudes and behavioural intentions (J. Kim et al., 2022; Obenza et al., 2024; Ramirez et al., 2024; Shamsi et al., 2022). However, there are still limited studies examining how students evaluate human trust, AI, and co-produced feedback, which this study aims to address.

**2.4. The Present Research**

This study fills these gaps by examining university students' perceptions of feedback generated by AI, humans, and their co-production, and by investigating changes in perception when the sources of feedback are revealed. The findings provide empirical evidence on which types of feedback should be prioritised. How they should be implemented effectively, promoting trust and the adoption of feedback in future HE contexts. It also aims to identify student characteristics that affect their ability to distinguish between feedback sources, which is vital for designing personalised and engaging feedback. We conducted a sequential investigation progressing from detection abilities to evaluative responses and underlying biases, offering insights essential for effective LLM integration in HE.

More specifically, the study addresses the following three interconnected research questions:

*RQ1. What factors predict students' ability to distinguish between AI-generated, human-generated, and co-produced feedback?*

This question establishes the baseline for students' detection ability, crucial for understanding the initial perceptions explored in RQ2.

*RQ2. To what extent does the students' perception of feedback change after its source has been revealed to them?*

Building on the findings of RQ1, this question examines whether source disclosure creates systematic shifts in perception, revealing the influence of provider identity versus content quality.

*RQ3. What biases emerge against the three types of feedback studied, and to what extent do these biases vary based on students' backgrounds?*

This question explores underlying mechanisms driving RQ2's perception changes, examining how student characteristics contribute to systematic preferences toward different feedback sources.

## 3. Methodology

This study employed a within-subject experimental design examining feedback identity disclosure effects, following the procedure outlined by Nazaretsky et al. (2024a). To investigate students' perceptions of different feedback providers, AI, human (TAs), and co-produced feedback (from both AI and humans), a multi-page web application was developed to deliver an online survey to students[1]. The web pages were designed to follow the experimental procedure (Figure 1).

The procedure included: (1) presenting study information and obtaining informed consent; (2) displaying students' weekly assignment answers alongside three feedback versions and seeking blind ratings on usefulness, objectivity, and genuineness; (3) measuring students' AI trust using the AI Trust Survey (Nazaretsky et al., 2025a); (4) introducing primer manipulation and requesting students identify AI and co-produced feedback conditions in a simplified Turing test (one attempt allowed); (5) revealing feedback origins (as either TA, AI or co-produced feedback) and asking students to re-evaluate their perceptions (usefulness, objectivity, genuineness) with new information, plus rating feedback provider credibility. Finally, demographic information was collected, including degrees, gender, age, and years of AI usage (in both general and educational contexts).

[1] The platform was built using the Svelte JavaScript framework for the front end and deployed on Amazon Web Services (AWS). The backend was implemented with FastAPI, while AWS DynamoDB was used for storing data in JSON format.

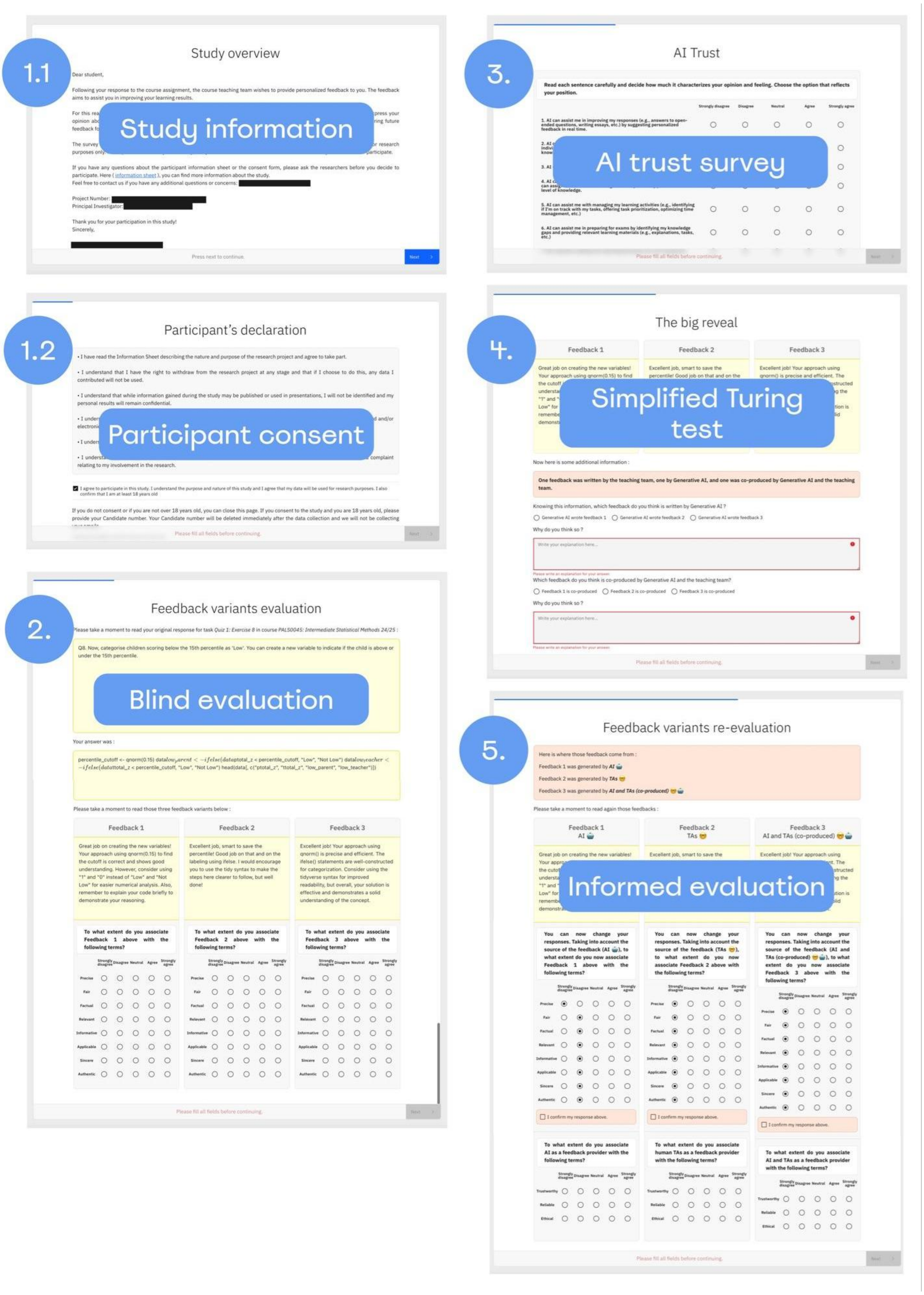

Figure 1. The experimental procedure.

### 3.1. Participants

An a priori power analysis was conducted to evaluate statistical power. The analysis for repeated measures ANOVA within factors assumed a medium effect size (f = 0.25), a significance level of 0.05, and anticipated three groups with two repeated measurements per participant. Accounting for a correlation among repeated measures of 0.5 and a nonsphericity correction of 1, the analysis yielded a numerator degree of freedom (df) of 1 and a denominator df of 79. Results indicated that approximately 82 participants would be needed to achieve power (1-β error probability) of 0.99, substantially exceeding the recommended 80% threshold. Based on this analysis, we aimed to recruit approximately 90 participants to ensure adequate statistical power to detect medium-sized effects using multilevel modeling with F tests, while minimizing Type II error risk.

Ninety-one first- and second-year psychology students from a UK University (blinded for review), enrolled in two statistical courses, participated in this five-week study. This study received institutional ethical approval (blinded for review) and a separate approval from the course leader to conduct experiments within their classes. Recruitment took place during students' practical sessions, where researchers introduced the study and provided incentives, including course credit and entry into a £50 Amazon voucher raffle. Online recruitment was conducted via email to targeted students and through course announcements. All participation was voluntary, and informed consent was obtained before the experiment.

Demographic data are presented in Table 1. Due to the low representation of the non-binary group (n = 2), these participants were combined with the majority female group (82%), resulting in two gender groups. Students with over four years of general AI usage (n = 2) were excluded from the analysis. Students also reported their age. Seven participants had missing information due to technical errors. Therefore, 82 responses were included.

Table 1. Demographic information of participants.

| **Course** | **Degree** | **Year** | **Number of participants(female/male/non-binary)** | **Age (SD)** | **Years of educational AI usage (SD)** |
|---|---|---|---|---|---|
| Intermediate Statistical Methods 24/25 | BSc, MSc | 2 | 27 (23/3/1) | 20.11 (2.10) | 1.33 (0.73) |
| Introduction to Statistical Methods 24/25 | BSc, MSc | 1 | 55 (45/9/1) | 18.60 (0.83) | 1.47 (1.02) |

### 3.2. Feedback Generation

For the experiment, assignments on chi-square analysis and factor analysis were selected for first-year and second-year students, respectively. Module leaders identified these questions as particularly challenging topics where students would benefit from detailed feedback. We coordinated with course leaders to ensure TAs provided substantive feedback (at least 2-3 lines) for these questions.

Following preliminary evaluation of LLM capabilities for educational feedback generation (detailed in Appendix A), Claude 3.5 Sonnet was selected as the primary model for this study based on its superior alignment with human teaching standards compared to GPT-4o. The evaluation revealed that Claude-generated feedback achieved 60% strong or complete alignment with TA feedback, significantly outperforming GPT-4o's 20% alignment rate.

After collecting TA feedback, we used a custom Python script utilising the Anthropic Claude API to generate AI and co-produced feedback variations, with all feedback generation conducted within connected sessions to maintain contextual consistency across feedback types. The script processed each student submission in conjunction with the question prompt and model answer to generate relevant AI feedback. For co-produced feedback, the script additionally incorporated original TA feedback, instructing the AI to evaluate and potentially improve human feedback while maintaining its core message.

During the experimental phase, three feedback variants were generated per student response:

1. **Human Feedback.** Feedback was provided by TAs, who routinely offer formative feedback on students' assignments. Before each course, teaching teams receive training, including guidelines, feedback examples, and quality moderation by course leaders. The average student-to-grader ratio is 30:1.
2. **AI feedback.** Claude 3.5 Sonnet API was utilised to create AI feedback via a prompt template (Appendix A). It was prompted to create formative feedback for specific tasks and students' responses. Given the formative focus and challenges with AI adherence to rubrics, no grading rubric was applied. Since statistical tasks are well-structured with single correct solutions, Claude was instructed to offer formative feedback while maintaining emotional support.
3. **Co-produced Feedback.** Human-AI co-produced feedback represented realistic scenarios where instructors utilised LLMs as AI tools to enhance feedback practices rather than replacing them entirely. For this variant, TAs' original feedback was submitted to Claude 3.5 Sonnet with instructions to evaluate and improve existing feedback while maintaining core pedagogical intent.

The prompt (Appendix A) instructed Claude to assess both student submissions and instructor feedback, providing improved versions only if improvement was deemed necessary. If the AI determined that the instructor's feedback did not need improvement, it would return the original feedback unchanged. In this study, all human feedback instances were modified by the AI. All experimental materials were prepared for each student before lectures, incorporating TA feedback. For all three feedback types, we assessed quality against model answers, which represented the benchmark for accuracy and completeness in addressing statistical tasks.

Figure 2 illustrates these three feedback variants as applied to a statistical task from student assignments.

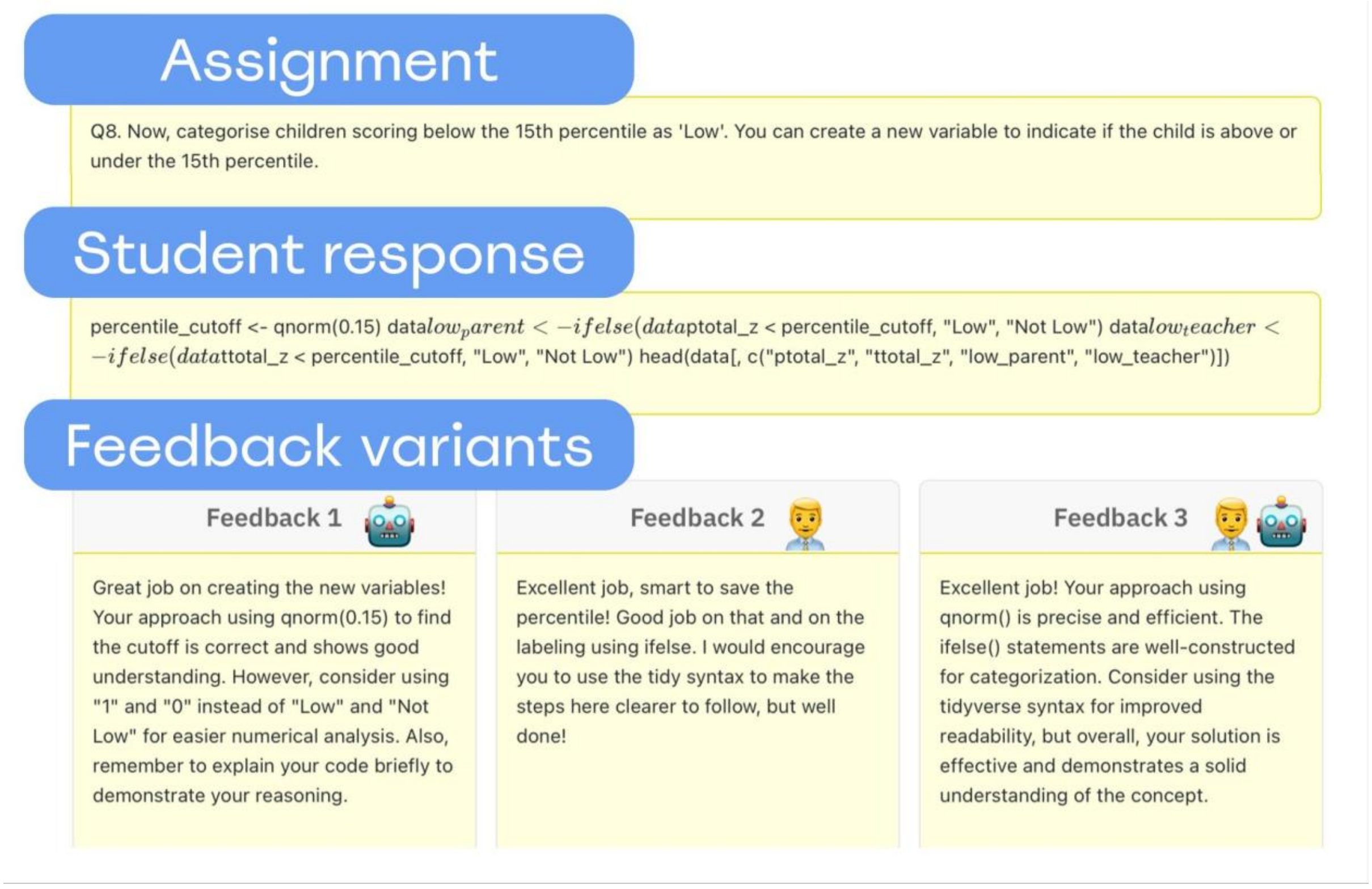

Figure 2. A sample of feedback variants.

### 3.3. Instruments

Following Nazaretsky et al. (2024a), students' perceptions of feedback content were measured using eight items that covered three dimensions: *Objectivity* (*Precise*, *Fair*, *Factual*), *Usefulness* (*Relevant*, *Informative*, *Applicable*), and *Genuineness* (*Authentic*, *Sincere*). Feedback sources were evaluated on *Credibility* using three terms: *Trustworthy*, *Reliable*, and *Ethical*. Both evaluations used single questions per term (Tormey, 2021), asking *"To what extent do you associate this feedback/feedback provider with the following term?"* Students rated terms using a 5-point Likert scale (1 = Strongly Disagree, 5 = Strongly Agree). All subscales demonstrated strong internal consistency, with Cronbach's alpha (α) values exceeding 0.88 for feedback content dimensions and 0.89 for feedback provider dimensions.

Additionally, Nazaretsky et al. (2025b) investigated students' trust in and adoption of AI using an instrument comprising 21 items on a 5-point Likert scale across four dimensions: *Perceived Usefulness* (6 items), *Perceived Obstacles* (7 items), *Perceived Readiness* (2 items), and *Trust Intent* (6 items). The instrument demonstrated strong internal consistency with Cronbach's alpha exceeding 0.75 across all dimensions.

### 3.4. Data Analysis

To answer RQ1, a logistic regression model was fitted to reveal correlations between demographics (age, gender, course), years of general AI usage, years of educational AI usage, the AI trust instrument, and students' ability to correctly guess feedback providers (Turing test results).

For RQ2, Mixed Linear Models (MLM) were constructed for each perception subscale (Objectivity, Usefulness, Genuineness) to reveal interactions between feedback provider (AI/Human/Co-produced) and timing (Blind/Informed), along with possible covariates (years of general AI usage, years of educational AI usage, AI trust instrument, age, gender, course). Intra-participant variation was modelled as a random effect.

Finally, a mixed linear model was developed to answer RQ3. The MLM included feedback provider (AI/Human/Co-produced) as a predictor, possible covariates (Turing test results, years of general AI usage, years of educational AI usage, AI trust instrument, age, gender, course), and a random effect (intra-individual variation between students). Tests of effect sizes were applied to significant correlations.

## 4. Results

We analysed 82 students' responses to the blind evaluation, simplified Turing Test (TT), and informed evaluations to investigate (1) factors predicting students' ability to distinguish AI, human, and co-produced feedback; (2) how students' perceptions changed when feedback provider identity was revealed; and (3) whether they hold biases against AI and co-produced feedback.

### 4.1. Factors Predicting AI Feedback Recognition

Of 82 students, 41 could not identify either AI or co-produced feedback correctly; 9 correctly identified AI feedback only; 9 correctly identified only co-produced feedback; and 23 correctly identified both AI and co-produced feedback.

Years of educational AI usage significantly predicted participants' ability to correctly identify AI-generated feedback ($\chi 2 = 4.08$, p = .028, Figure 3), but did not significantly influence their ability to identify co-produced feedback ($\chi 2 = 0.64$, p = .802, Figure 3). Years of general AI usage, age, gender, course, and AI trust showed no significant correlations.

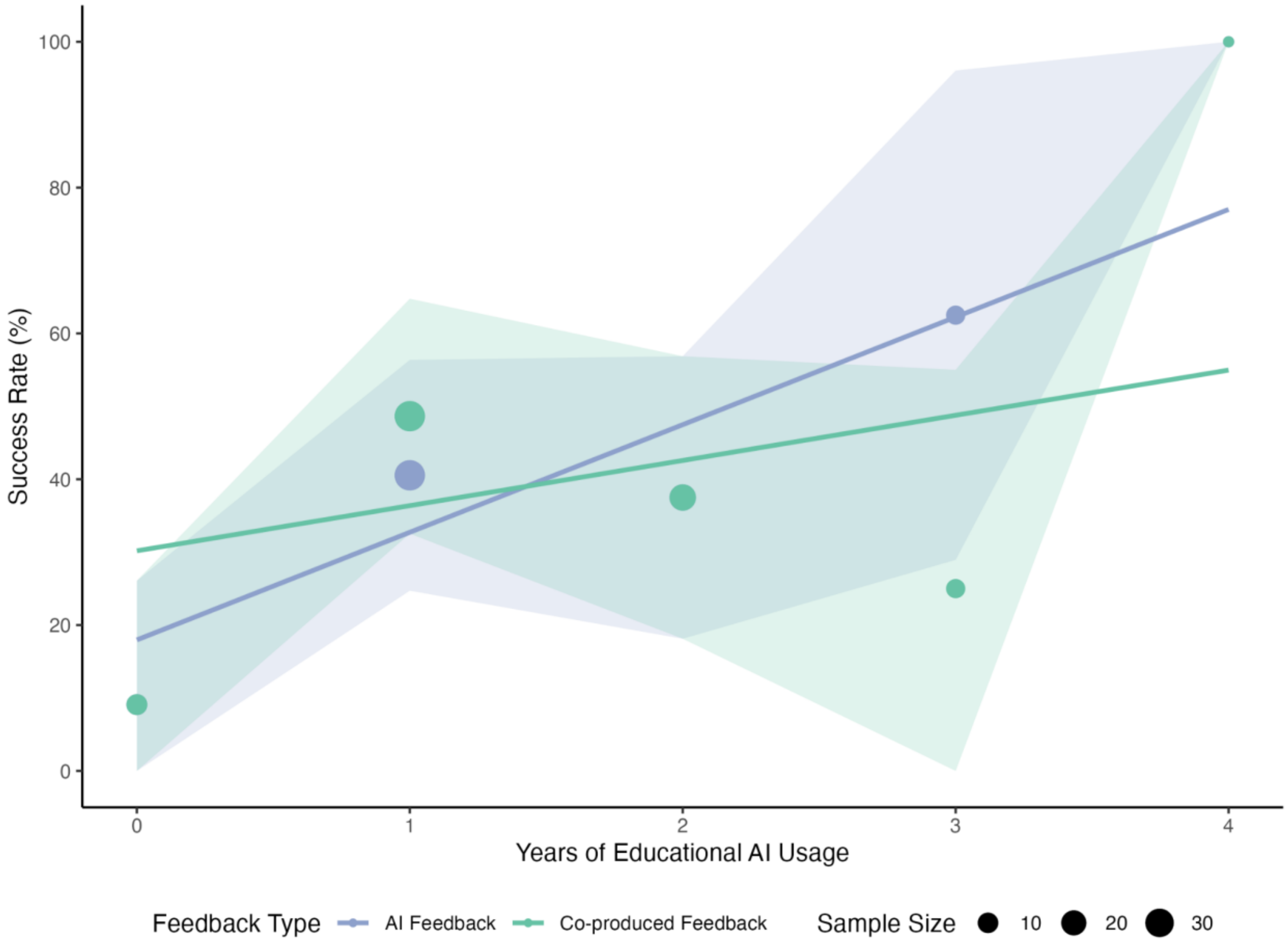

Figure 3. Success rates of AI feedback identification versus co-produced feedback identification across years of educational AI usage (0-4). AI feedback identification showed significantly stronger improvement with educational AI experience ($p = .028$). Shaded areas represent confidence intervals; point sizes indicate relative sample sizes.

## 4.2. Perception of Different Feedback Types

Feedback providers had a significant main effect on students' perceptions. AI feedback was rated significantly higher than human feedback across all dimensions (all $p < .05$). Similarly, co-produced feedback was rated significantly higher than human feedback across all dimensions (all $p < .001$). AI and co-produced feedback were rated similarly across all dimensions.

For the Objectivity dimension, AI ($M_{Blind}$ = 4.11, $M_{Informed}$ = 4.01) and Co-produced feedback ($M_{Blind}$ = 4.07, $M_{Informed}$ = 4.08) were rated similarly, both significantly higher than Human feedback ($M_{Blind}$ = 3.31, $M_{Informed}$ = 3.39, $p < .001$). Usefulness showed similar patterns, with AI ($M_{Blind}$ =4.12, $M_{Informed}$ = 4.03) and Co-produced feedback ($M_{Blind}$ = 4.03, $M_{Informed}$ = 3.99) rated similarly, both significantly higher than Human feedback ($M_{Blind}$ = 3.17, $M_{Informed}$ = 3.25, $p < .001$). For Genuineness, the AI feedback ($M_{Blind}$ = 4.11, $M_{Informed}$ = 3.67) was rated similarly to Co-produced feedback ($M_{Blind}$ = 3.82, $M_{Informed}$ = 3.84 ), both significantly higher than Human feedback ($M_{Blind}$ = 3.19, $M_{Informed}$ = 3.32, $p < .05$).

The main effect of timing (Blind/Informed) was only significant in the Genuineness dimension ($p < .001$), not for Objectivity and Usefulness (Figure 4). Subsequent Tukey-adjusted mean comparisons for Genuineness revealed a significant feedback provider × timing interaction: AI feedback was rated as more genuine in the Blind condition than in the Informed condition ($p < .001$), whereas co-produced and human feedback remained stable across timing conditions.

Two covariates significantly predicted Usefulness ratings but not Objectivity or Genuineness (Figure 5). More years of general AI usage were associated with lower Usefulness ratings ($p = .024$, $\beta = -0.36$, Figure 5, left). Male participants rated feedback significantly lower than female and non-binary participants regarding Usefulness ($p = .051$, $\beta = -0.40$, Figure 5, right). Subsequent Tukey-adjusted mean comparisons revealed a significant interaction between feedback provider and gender: human feedback was rated significantly less useful by male students ($M$ =

2.67) than by female and non-binary students (*M* = 3.46, *p* = .001). We found no significant effects of Turing test results, course, age, or AI trust instrument on feedback perception.

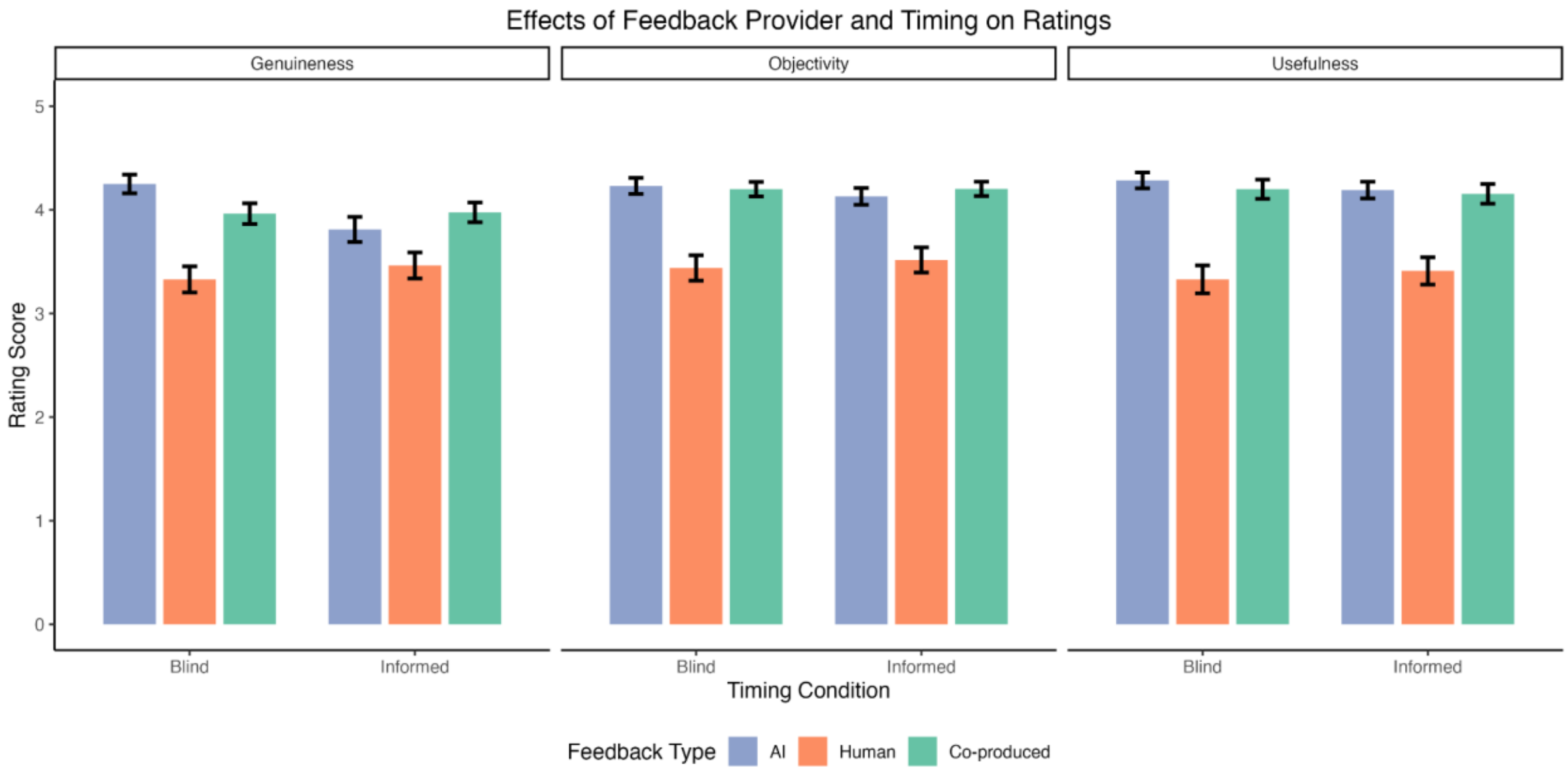

Figure 4. Perceptions towards feedback content per dimension (Genuineness, Objectivity, Usefulness) by feedback provider (AI/Human/Co-produced) and Timing (Blind/Informed) with error bars.

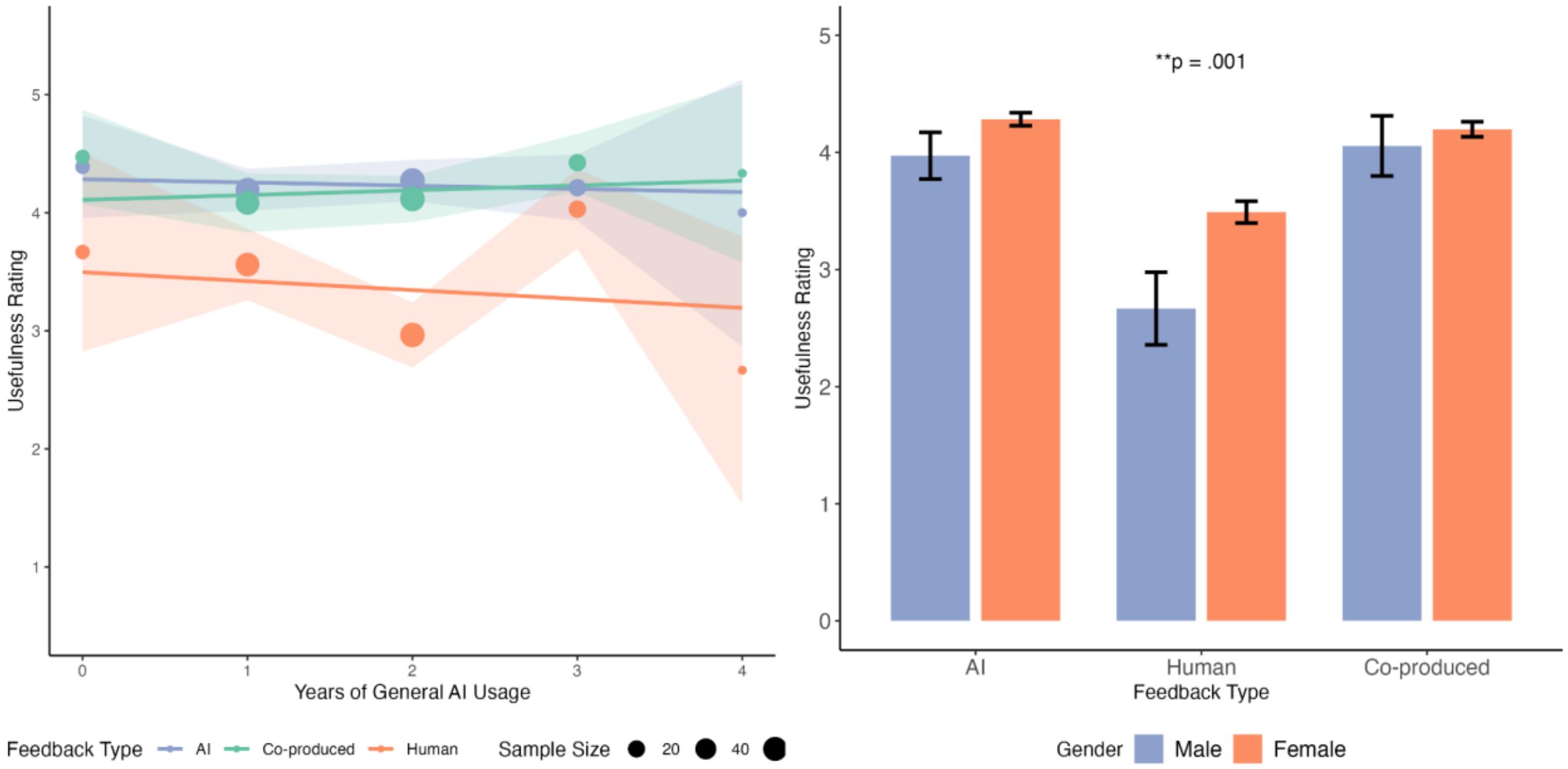

Figure 5. Years of General AI Usage and Gender Differences in Feedback Usefulness Perceptions. The left panel shows usefulness ratings by years of general AI usage across three feedback types (AI, Human, Co-produced), with confidence intervals (shaded areas) and trend lines. More general AI experience correlated with lower Usefulness ratings ($p$ = .024). The right panel displays gender differences in perceived usefulness by feedback type with standard error bars. Males rated Usefulness significantly lower than females for human feedback ($p$ = .001). Scores ranged from 1 (Strongly Disagree) to 5 (Strongly Agree).

**4.3. Bias against AI and Co-produced Feedback**

Participants rated Human feedback ($M$ = 4.18, $SD$ = 0.84) significantly more credible than AI feedback ($M$ = 3.28, $SD$ = 0.84) with a large effect size ($p$ < .001, Cohen's $d$ = 1.21, Figure 6a). Co-produced feedback ($M$ = 4.04, $SD$ = 0.69) was also rated significantly more credible than AI feedback with a substantial effect ($p$ < .001, Cohen's $d$ = 1.02, Figure 6a). Credibility ratings were similar between Human and Co-produced feedback.

Three covariates significantly predicted credibility ratings. More years of general AI usage led to lower credibility ratings ($p$ = .025, $\beta$ = -0.27), with no significant feedback provider × years of general AI usage interaction (Figure 6b). In contrast, students who used educational AI more rated feedback as more credible ($p$ = .027, $\beta$ = 0.27), with no significant interaction between feedback provider and years of educational AI usage (Figure 6b). Finally, perceived usefulness in the AI trust instrument (usefulness_AITrust) significantly predicted credibility ratings ($p$ = .031), with significant feedback provider × perceived usefulness interaction: students with higher perceived AI usefulness ratings gave significantly higher credibility ratings to AI-generated feedback specifically ($p$ = .019, β = 0.30), while this relationship was not significant for human-generated ($p$ = .549, β = 0.09) or co-produced feedback ($p$ = .236, β = 0.14, Figure 6c). Other AI trust instrument dimensions (perceived obstacles, perceived readiness, trust intent) showed no significant effects on credibility ratings. Additionally, Turing test results, age, gender, and course variables showed no significant effects.

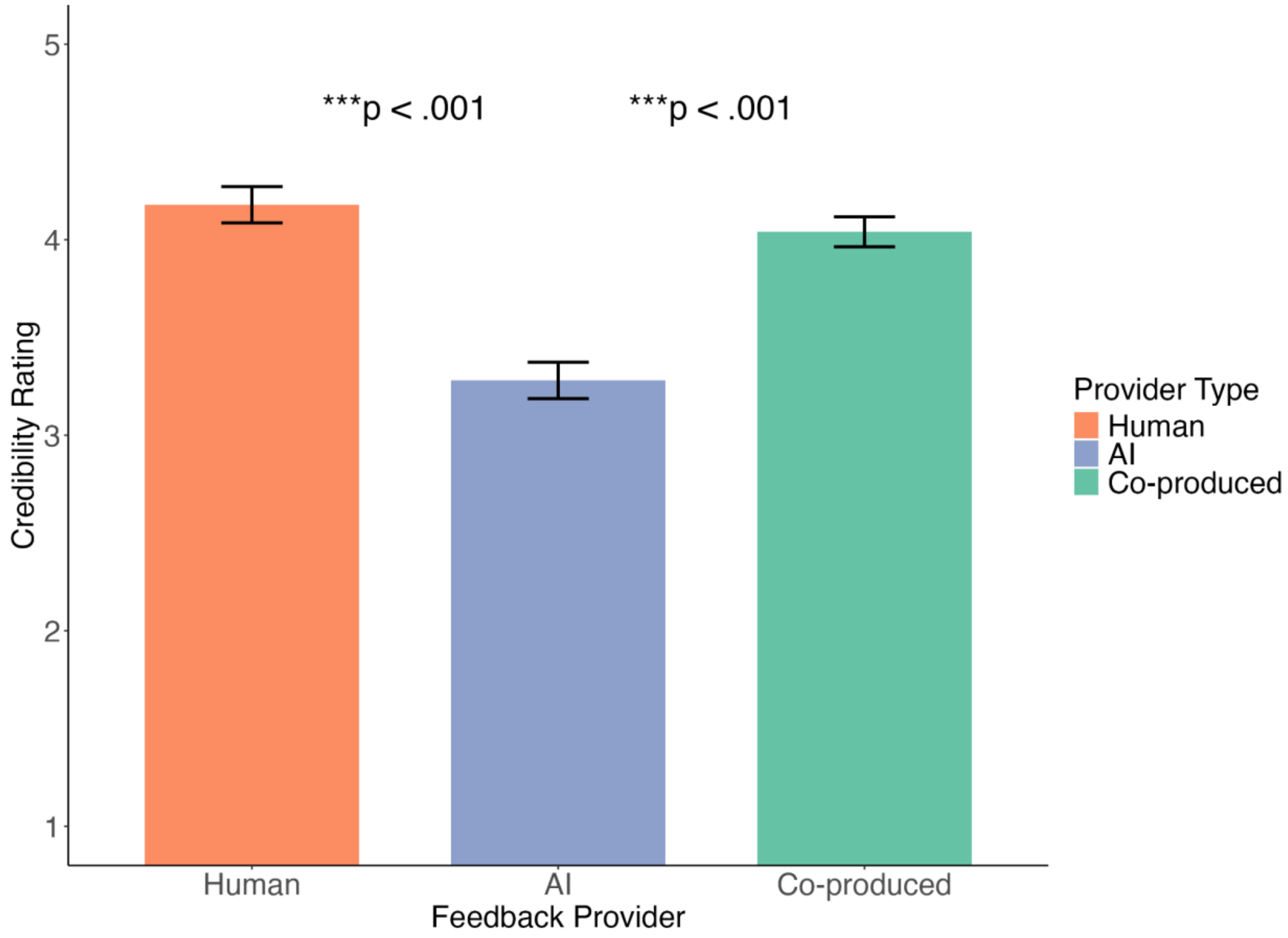

Figure 6. a

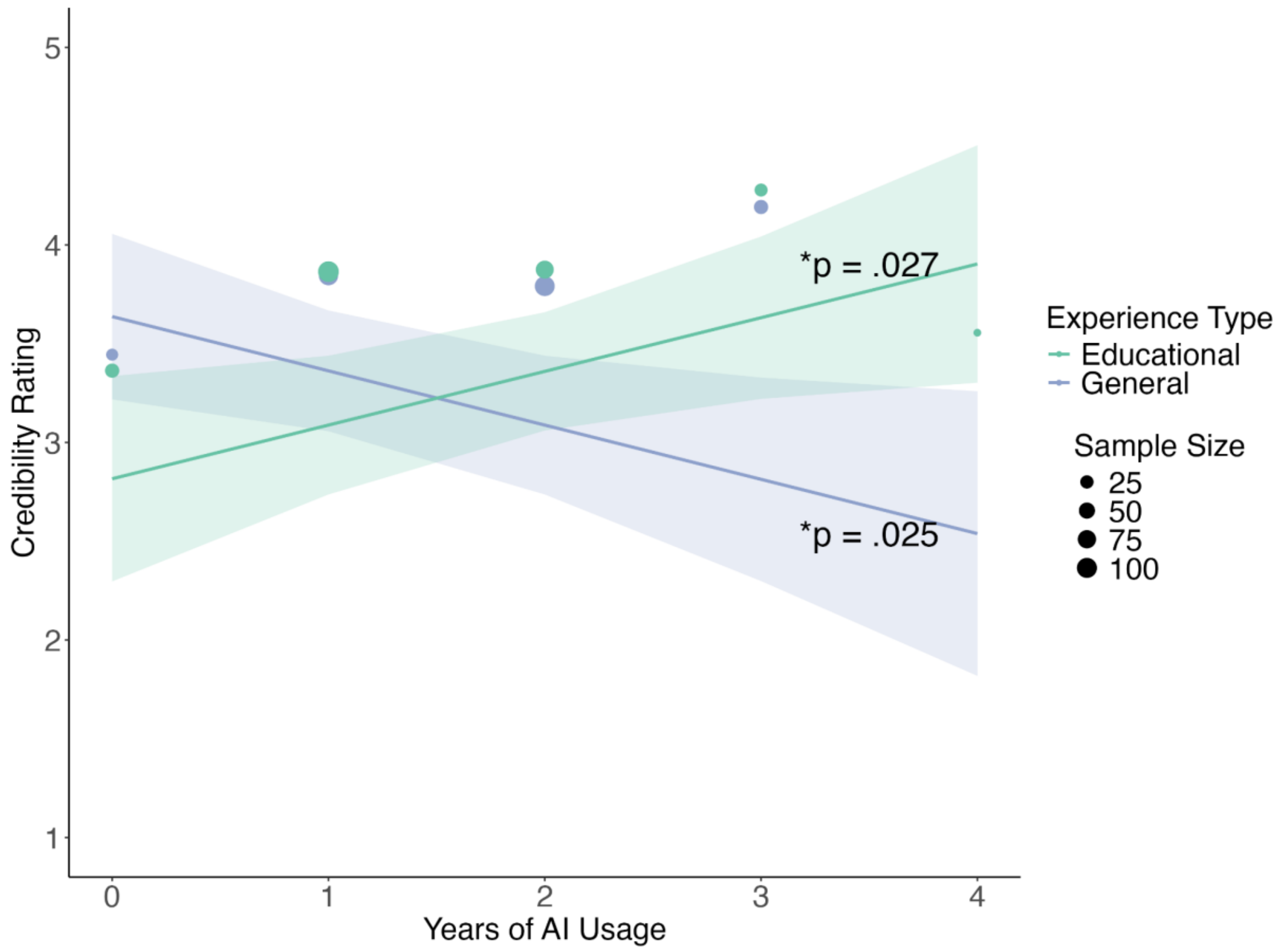

Credibility Rating
Years of AI Usage
*p = .027
*p = .025
Experience Type
Educational
General
Sample Size
25
50
75
100

Figure 6. b

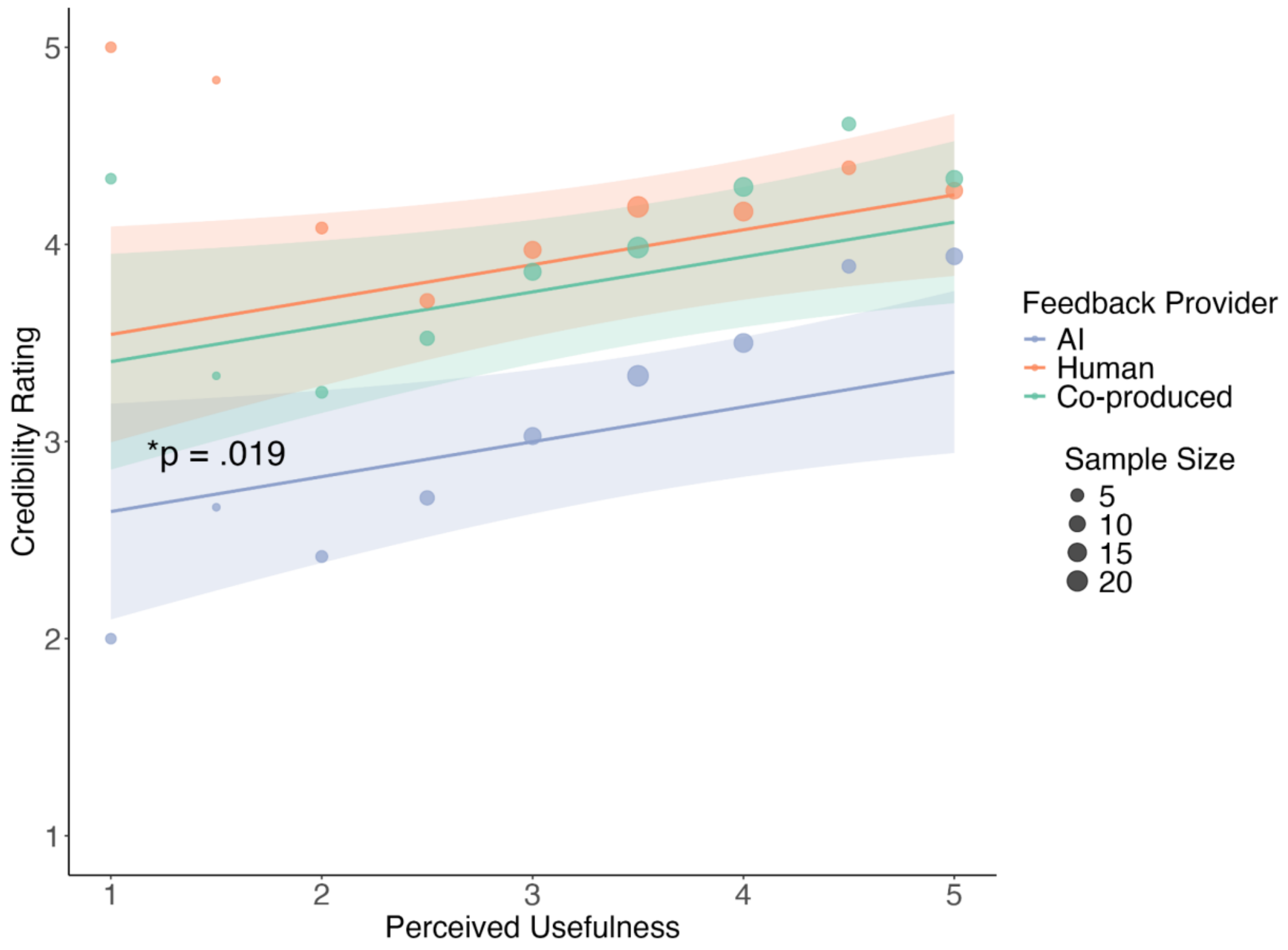

Figure 6. c

Figure 6. Credibility Perceptions by Feedback Provider, AI Experience, and Perceived Benefits. (a) Mean credibility ratings by feedback provider type, with Human and Co-produced feedback rated significantly higher than AI feedback ($p < .001$). (b) The relationship between years of AI usage and credibility ratings showed opposite effects for educational and general AI experience types. More years of general AI usage predicted lower credibility ratings ($p = .025$), while more years of educational AI usage predicted higher ratings ($p = .027$). Shaded areas represent 95% confidence intervals. Sample sizes were indicated by point size. (c) Influence of perceived usefulness on credibility ratings across three feedback types, with AI feedback showing a significant positive correlation ($p = .019$). Sample sizes are indicated by point size. All ratings ranged from 1 (Strongly Disagree) to 5 (Strongly Agree).

## 5. Discussion

This study revealed three key findings about student perceptions of feedback across different provider types. First, students consistently rated both AI-generated and co-produced feedback significantly higher than human feedback on objectivity and usefulness dimensions, while credibility perceptions favored human and co-produced feedback over AI-only feedback. Second, educational AI experience emerged as a crucial factor, significantly predicting students' ability to identify AI-generated feedback and correlating with higher credibility ratings across all feedback types. Third, disclosure of feedback source primarily affected perceptions of genuineness, with AI feedback experiencing the largest decline, while functional dimensions like usefulness and objectivity remained unaffected.

These findings can inform real-world LLM use cases in HE contexts where teaching teams resist AI feedback, debate full AI adoption, or experiment with AI-enhanced human feedback (Holstein & Aleven, 2022; Pahi et al., 2024). Below, we discuss the findings in more detail.

### 5.1. Students’ Ability to Distinguish AI, Human, and Co-produced Feedback

Results revealed students' differentiated ability to distinguish between AI-generated and co-produced feedback. Specifically, more years of educational AI usage correlated with better identification of purely AI-generated feedback, but not co-produced feedback. This differential recognition likely reflects two factors: co-produced feedback is inherently harder to identify, as it requires distinguishing it from purely human feedback, and students likely have limited exposure to such systems since no widely adopted educational applications currently implement this hybrid approach. Without prior experience with co-produced feedback, students lack the reference points necessary to develop recognition skills, regardless of their general educational AI experience. This "educational AI usage" factor, newly introduced in our study and not previously examined in the literature (Nazaretsky et al., 2024a), suggests that as students increasingly incorporate AI into education, they develop greater familiarity with the distinctive content features generated by LLMs.While established frameworks like TAM (Technology Acceptance Model) and MDMT (Multi-Dimensional Measure of Trust) have long measured constructs such as "usefulness" and "benefits" in technology adoption, our study added educational AI trust survey items to capture the unique aspects of AI integration in learning contexts, distinguishing educational AI experience from general AI usage patterns.

Consistent with Nazaretsky et al. (2024a), no additional factors significantly predicted Turing test performance. Notably, while Nazaretsky et al. (2024a) reported 60% of participants successfully identified AI feedback in a two-condition study (human vs. AI), our three-condition experiment revealed distinctive results: only 40% correctly identified AI-generated feedback. This decreased identification accuracy suggests that introducing co-produced feedback may have increased cognitive load, potentially compromising students' ability to distinguish between human and non-human feedback.

### 5.2. Perception of Different Feedback Types

Students strongly preferred AI-generated feedback and co-produced feedback over human feedback, contradicting Nazaretsky et al. (2024a), who found that human feedback was preferred over AI-generated alternatives. Similar to Le et al. (2025)'s study, Chinese university students (N=114) demonstrated a strong preference for human feedback (87.2%) over ChatGPT (4.0) generated feedback. However, other studies support AI preference patterns. In English as a Foreign Language contexts, Escalante et al. (2023) found that students preferred ChatGPT-generated feedback over human tutor feedback in a 6-week longitudinal study, with participants rating AI feedback higher on satisfaction, comprehensibility, helpfulness, and overall preference dimensions.

One explanation for students' preference patterns may lie in the contextual factors of our study. Our research was framed within a real-world teaching context, where a TA graded 30 student assignments weekly, potentially compromising the quality of human feedback. This aligns with the findings of Dai et al. (2023), who suggested that AI feedback may demonstrate superior readability, detail, and accuracy compared to human feedback under such time-constrained conditions. Henderson et al. (2025) provide additional context through their large-scale cross-disciplinary survey, which identified that AI and teacher feedback serve complementary rather than competing functions: AI feedback was perceived as more accessible, understandable, and less risky, while teacher feedback was seen as more credible, insightful, relational, and contextualised. Further research is needed to identify the reasons behind students' preferences for different feedback types, which are currently out of the scope of the current study. However, our results indicate that AI-generated feedback might be limited regarding students' adoption and sustained engagement in the long term due to students' bias against AI as a source of feedback in their formal HE.

Surprisingly, students did not prefer co-produced feedback over AI feedback, suggesting that the qualities this hybrid approach aimed to enhance did not necessarily align with students' actual needs, reflected in their lower ratings of usefulness, objectivity, and genuineness. This unexpected finding may be attributed to the generic prompts deployed for co-produced feedback. Unspecified criteria for LLM improvement could lead to enhancements that are generic and might not align with students' valued criteria, such as motivational support and clarifying examples (Pahi et al., 2024), potentially making the hybrid approach indistinguishable from purely AI feedback.

Compared to previous literature, students rated AI-generated feedback as less genuine once they were informed of its source. Perceptions of usefulness and objectivity remained unchanged after the feedback identity was revealed. Unlike usefulness and objectivity, genuineness engages more directly with affective aspects of AI adoption. This isolated decline in perceived genuineness might reflect algorithmic aversion, where negative attitudes toward algorithms influence responses to AI systems (Dietvorst et al., 2014). Our findings highlight the emotional resistance that learners may exhibit toward the integration of LLMs in education.

Notably, students did not adjust ratings when the co-produced feedback identity was revealed. Learners appeared to discount the AI contribution in co-produced feedback, continuing to perceive it as more genuine, helpful, and objective than purely human feedback. This finding contributes to understanding how nuances of AI framing can impact learners' perceptions of educational quality (Cukurova et al., 2020): when educational evidence is purely presented within an AI research framework, nontechnical audiences, including teachers and students, tend to consider it less prestigious and rigorous than identical evidence framed within traditional disciplines. Compared to teachers’ reluctance to embrace AI in feedback generation (Kizilcec, 2024), both our research and Nazaretsky et al. (2024a) suggest learners hold more nuanced attitudes toward varying degrees of AI involvement in feedback.

### 5.3. Biases Against AI and Co-produced Feedback

Students demonstrated bias against AI-generated feedback, showing significantly lower credibility ratings compared to human feedback. These bias patterns align with broader algorithm aversion research, which demonstrates that humans consistently prefer human judgment over algorithmic decisions, even when algorithms demonstrably outperform humans (Dietvorst et al., 2015; Burton et al., 2020). However, students did not extend this bias to co-produced feedback, rating its credibility similarly to human-generated feedback. This distinction illustrates that when feedback is framed as a collaborative human-AI effort, learners tend to trust that it maintains the same quality and credibility as purely human feedback. These findings suggest the benefits of the hybrid approach in the context of feedback provision (Cukurova, 2019, 2024), not only regarding the content of feedback where both human and AI agents can combine their complementary strengths to achieve outcomes greater than the individual agents' alone, but also for better alignment with student perceptions and expectations for feedback in HE contexts. This offers a pragmatic approach for educators in implementing co-produced feedback (explicitly labelled as such) as an effective middle ground between fully automated and human methods, balancing AI benefits in generating detailed feedback, similar to (Escalante et al., 2023; Luo et al., 2025), while maintaining positive student perceptions potentially attributing to teacher involvement (Er et al., 2024).

### 5.4. Individual Differences in Identifying and Trusting AI

The emergence of general AI usage and educational AI usage as significant new factors in our study was notable. Both AI experience variables significantly predicted AI feedback recognition and perception. Students with more educational AI usage were better at identifying AI-generated feedback. Research shows that AI and human-generated content display distinctive linguistic patterns across dimensions, including involvement versus informational production, narrative versus non-narrative concerns, explicit versus context-dependent references, persuasive expression, and abstract versus concrete style (Biber, 1991; Sardinha, 2024). While we controlled for specific features by instructing the AI to use second-person pronouns and similar length constraints as human feedback, students with greater educational AI experience likely detected subtler linguistic markers in the AI feedback, enabling them to accurately identify its non-human origin. Co-produced feedback potentially retained human-generated content as its foundation, incorporating natural linguistic patterns that made it considerably harder to distinguish from purely human-authored feedback.

This finding aligns with the results of RQ3, where increased educational AI usage predicted higher credibility ratings for both AI-generated and co-produced feedback. The more students used AI in educational contexts, the more they trusted it. Surprisingly, general AI usage showed an opposite pattern: students with more experience using general AI rated AI and co-produced feedback as less useful and less credible. While this divergence is noteworthy, it should be interpreted with caution, as it is based on a single questionnaire item. Nevertheless, this trend suggests that different types of AI exposure may have different effects on students' trust.

This pattern of experience-dependent trust reflects the importance of the broader development of student AI literacy in educational contexts. Students who regularly use AI tools for tasks such as explaining concepts, summarising articles, and searching for information tend to view these systems as enhanced study tools rather than autonomous agents (Common Sense Media et al., 2024; Freeman, 2025). This practical, tool-based perspective builds confidence in AI capabilities over time. Zou et al. (2025) emphasized the importance of examining how students' AI literacy, defined as 'knowledge and skills regarding using AI tools for writing', might influence their acceptance of ChatGPT-generated feedback in writing contexts. Our findings support this connection between familiarity and trust: students who scored higher on the perceived usefulness of AI also gave higher credibility ratings to AI feedback. This suggests that when students understand AI as a sophisticated but controllable educational resource, similar to advanced search engines or proofreading websites, they become more willing to trust and engage with AI-generated content.

These results offer two key insights: (1) The more students used AI in learning contexts, the more they trust AI feedback; (2) Greater general AI usage may lead students to be more critical of AI in educational contexts, possibly explained by mixed effects of AI feedback, helpful with basic errors but potentially disruptive to self-regulated learning in complex problem-solving scenarios (Vanacore et al., 2025). While previous research has established that student trust in AI increases the perceived usefulness of AI educational technology (Nazaretsky et al., 2025a), our findings suggest that this relationship works bidirectionally. The contrasting effects between general and educational AI usage indicate that students' heightened trust in AI feedback may reflect unjustified confidence rather than warranted trust. This highlights the importance of developing students' critical evaluation skills, particularly with AI applications in educational settings (Laupichler et al., 2022; Walter, 2024).

Research demonstrates that students require structured guidance to develop AI literacy competencies, as many exhibit concerning patterns of uncritical acceptance of AI-generated content (Walter, 2024; Zhang & Magerko, 2025). The AI literacy framework by Long and Magerko (2020) emphasizes that AI literacy fundamentally involves the ability to "critically evaluate AI technologies," a principle reinforced across multiple frameworks, including UNESCO's AI Competency Framework for Students (Miao & Shiohira, 2024) and Digital Promise's AI Literacy Framework (Mills et al., 2024). Students require structured guidance to develop a nuanced understanding of when and how to appropriately rely on AI-generated educational content, as multiple reviews and surveys indicate that excessive reliance on AI tools can negatively impact students' independent thinking and critical evaluation skills. For example, recent studies report that a large majority of students are concerned that using AI without proper supervision may limit their ability to think critically and make responsible decisions (Szmyd & Mitera, 2024; Sardi et al., 2025; Melisa et al., 2025).

## 6. Recommendations for Practitioners

Instructors facing resource constraints, particularly in large-enrollment courses, should consider AI-assisted feedback as a strategic supplement to human expertise rather than a replacement. The shift to AI-assisted feedback necessitates developing new competencies in AI instruction, leveraging contextual understanding to tailor and evaluate feedback for meaningful educational experiences (Nazaretsky, 2025). Instructors should emphasize AI's practical benefits, enhanced consistency, speed, and comprehensiveness when introducing these systems, as the stability of co-produced feedback ratings after disclosure suggests that partial AI involvement is more acceptable to students than full automation. Transparency remains crucial: instructors must clearly communicate when and how LLM tools are used in feedback provision, allowing for gradual introduction through co-production models while maintaining human oversight.

Students should be educated to recognize that AI-assisted feedback serves different functions than purely human feedback, with each offering distinct educational value. Those new to LLM tools benefit from structured introductions that emphasize practical benefits and provide explicit guidance on how to engage effectively with AI-assisted content, while students with general AI experience should understand how educational applications differ from entertainment uses and adapt their engagement strategies accordingly. When receiving feedback that involves AI assistance, students might focus on the functional value, consistency, comprehensiveness, and detailed examples, while seeking human input for personalized guidance and complex conceptual issues. Understanding the hybrid nature of modern feedback systems can enable students to utilize feedback effectively regardless of its source, maximizing learning outcomes in AI-augmented educational environments.

## 7. Limitations and Future Research

Several limitations should be acknowledged. First, the study was conducted with a limited sample within a specific statistical course, hindering generalizability to other disciplines. Future studies should investigate the effectiveness of AI feedback in various academic contexts using larger samples to enhance the robustness of the findings.

Second, the design of AI-generated and human-AI co-produced feedback was constrained by the limitations of the LLM model, predefined prompts, and methodological approach. While this reflected real-world AI usage, other human-AI collaboration methods in feedback generation exist and warrant investigation. Additionally, the variation in experience and feedback styles among multiple TAs was not taken into account, which may affect the quality and consistency of feedback. Future research should examine how different teaching experience levels interact with AI-assisted feedback generation and how institutional training might standardise co-production practices across diverse teaching teams.

Third, the study relied on specific instruments to measure students' perceptions, potentially overlooking other influential factors such as learning performance. Future research should expand measurement scope by incorporating additional instruments, demographic characteristics, and performance-based analyses. Furthermore, qualitative follow-up interviews could provide deeper insights into students' trust or scepticism toward AI-generated and co-produced feedback.

Finally, the study employed logistic regression and mixed linear models to examine relationships between feedback provider identity, student demographics, and perceptions. While suitable for addressing the research questions, these models assume linear relationships between predictors and outcomes, which may not fully capture the complex dynamics of learning and perception. Nonlinear models, machine learning techniques, or Bayesian approaches could be explored to provide a more comprehensive understanding of AI feedback in educational contexts.

## 8. Conclusion

Students' contrasting perceptions of AI-generated, human-generated, and co-produced feedback highlight both opportunities and challenges in integrating LLM-powered AI into feedback practices in HE contexts. While students generally rated AI and co-produced feedback favourably regarding usefulness and objectivity, the influence of AI experience, gender differences, and algorithmic aversion suggests the need for thoughtful implementation strategies that balance technological capabilities with human expertise. Our findings contribute to research on the application of AI-generated feedback in HE contexts and provide practical insights for HE teachers and institutions seeking evidence-based guidelines for incorporating AI into feedback systems.

## List of Abbreviations

AI - Artificial Intelligence

API - Application Programming Interface

FPQ - Feedback Perceptions Questionnaire

GenAI - Generative Artificial Intelligence

HE - Higher Education

LLM - Large Language Model

METI - Muenster Epistemic Trustworthiness Inventory

MLM - Mixed Linear Models

RQ - Research Question

STEM - Science, Technology, Engineering, and Mathematics

TA - Teaching Assistant

TAM - Technology Acceptance Model

TT - Turing Test

UK - United Kingdom

## Declarations

### Availability of data and materials

The data that support the findings of this study are available from the corresponding author upon reasonable request.

### Funding

This research received no specific grant from any funding agency in the public, commercial, or not-for-profit sectors.

### Acknowledgements

The authors would like to thank the students who participated in this study and the course leaders who facilitated data collection. We also acknowledge the teaching assistants who provided the human feedback used in this research.